# Investigating Circularity in India's Textile Industry: Overcoming Challenges and Leveraging Digitization for Growth

Suman Kumar Das
University of Brescia
s.das001@unibs.it

## Abstract:

India's growing population and economy have significantly increased the demand and consumption of natural resources. As a result, the potential benefits of transitioning to a circular economic model have been extensively discussed and debated among various Indian stakeholders, including policymakers, industry leaders, and environmental advocates. Despite the numerous initiatives, policies, and transnational strategic partnerships of the Indian government, most small and medium enterprises in India face significant challenges in implementing circular economy practices. This is due to the lack of a clear pathway to measure the current state of the circular economy in Indian industries and the absence of a framework to address these challenges. This paper examines the circularity of the 93-textile industry in India using the C-Readiness Tool. The analysis comprehensively identified 9 categories with 34 barriers to adopting circular economy principles in the textile sector through a narrative literature review. The identified barriers were further compared against the findings from a C-readiness tool assessment, which revealed prominent challenges related to supply chain coordination, consumer engagement, and regulatory compliance within the industry's circularity efforts. In response to these challenges, the article proposes a strategic roadmap that leverages digital technologies to drive the textile industry towards a more sustainable and resilient industrial model.

## Introduction:

In recent years, the "Make in India" initiative has catalysed the growth and development of the small and medium enterprise sector in India (Chhimwal et al., 2022). The Indian Micro, Small, and Medium Enterprise (SMEs) sector, comprising 48 million enterprises, is a significant contributor to India's economy, employing 106 million people and accounting for 6.11% of the manufacturing GDP and 24.63% of the service sector GDP (Nudurupati et al., 2022). Whereas India's resource extraction is high at 1,580 tonnes per acre, which is 251% above the global average of 450 tonnes per acre (Sarma et al., 2023). Additionally, SMEs account for approximately 25% of the total energy consumption within the industrial sector. Furthermore, India is the third-largest emitter of greenhouse gases, contributing 9.2% of total global emissions (Varun Boralkar, 2023). Due to the high resource extraction and greenhouse gas emissions in India, environmental sustainability has become a critical concern. The collective responsibility of SMEs is of utmost significance in addressing this issue. In this context of sustainability, India is making strides toward adopting a Circular Economy(CE) to meet the country's Nationally Determined Contribution (NDC) targets (Ellen MacArthur Foundation, 2016; Gedam et al., 2021; Varun Boralkar, 2023). This involves the careful alignment and management of resource flows across the value chain, integrating forward and reverse logistics, design innovation, collaborative ecosystems, and the development of new business models. Despite India's positive initiatives like the Swachh Bharat Mission and Mission LiFE, which reflect the country's commitment to sustainable and CE practices, the adoption of CE practices by SMEs has not been widely implemented (Varun Boralkar, 2023). Several key challenges hinder the implementation of CE practices, such as technological gaps, insufficient data to inform strategic decisions, a lack of clarity and coherence in the regulatory framework, and a dearth of collaborative efforts across different sectors, including education, policy support, technological innovation, and enhanced consumer awareness (Das et al., 2024). Having sector-specific CE data is crucial for overcoming these barriers and improving the adoption of CE practices in India (Badhotiya et al., 2022; Bag et al., 2022; Nudurupati et al., 2022). These can help in design tailored strategic decisions framework for different sectors based on their unique characteristics, challenges, and opportunities. These sector-specific data facilitate the development of targeted strategies that address the specific needs and dynamics of each industry. It also provides benchmarks that enable companies to evaluate their performance against industry norms and best practices, thereby facilitating more effective implementation of CE principles.

This article examines the circular readiness of Indian textile industry. Over the past 15 years, India's textile sector has doubled its apparel production, making the country the 6th largest global exporter of textiles and apparel with promising growth prospects (Kanupriya, 2021). India's textile industry exemplifies regional specialization, with each state playing a pivotal role. Karnataka, cantered around Bengaluru and Mysuru, is celebrated for silk sarees and cotton textiles, contributing to both domestic and export markets. Tamil Nadu excels in handloom textiles, knitwear, and cotton fabrics, with Tirupur driving India's knitwear exports and cities like Coimbatore and Kanchipuram enriching its textile heritage. In Gujarat, cities like Surat, Ahmedabad, and Vapi produce 30% of India's cotton output, with Surat alone generating 30 million meters of raw materials daily. Rajasthan, known for its traditional block printing and handloom crafts, thrives in cities such as Jaipur, Ajmer, and Bhil. Maharashtra stands out for power loom and affordable fabrics, with Bhiwandi acting as a major supplier to garment manufacturers. Uttar Pradesh, home to the iconic Banarasi sarees from Varanasi, excels in luxurious silk products. Punjab, dominated by Ludhiana and Amritsar, leads in Woolen textiles and hosiery, meeting winter wear demands nationwide. West Bengal, with Kolkata and Murshidabad at the forefront, produces 75% of the global jute output while maintaining excellence in silk production (Ankita Sharma, 2020; CII, 2023; Sanjay Bakshi, 2024).

However, the textile industry is highly polluting, with excessive use of toxic dyes, water, chemicals, and auxiliary materials. India generates approximately 7.7 million tons of textile waste annually (Dwivedi et al., 2023; Ponnambalam et al., 2023). Although recycling textile waste at both pre- and post-consumer stages offers substantial opportunities, the current recycling rate remains below 1% (CII, 2023; Ncaer, 2009). The common challenges faced by these industries that limit the progression of CE practices are multifaceted and complex. These include the lack of adequate infrastructure and technological capabilities, the dearth of relevant methods, tools, and techniques to support CE implementation, the inherent complexities of product configurations, the lack of top management commitment and holistic vision towards sustainability. Ponnambalam et al., (2023) identified and analysed multiple barriers to implementing textile recycling in India. While various articles have discussed barriers to implementing CE practices, but a structured framework to help companies assess their current practices against CE principles is currently lacking. As a result, many companies struggle to determine where to start in their transition to a CE. The report by Dwivedi et al., (2023) analysed the readiness, but it lacks clear metrics and provides a broad evaluation without considering the various elements of circularity across different areas such as design, production, and end-of-life management. This results in an ambiguous response on how to implement CE principles in industrial organizations to achieve net-zero emissions. Therefore, it is crucial to measure circularity at the micro-level to prioritize targeted actions for improving circular practices. Hence, this article aims to answer the following research gap.

- ***How circular is the Indian textile industry?***
- ***What barriers need to be addressed to improve circularity in textile industry?***

To evaluate the circularity of the manufacturing companies, we used the C-readiness Tool developed by the RISE lab at the University of Brescia (Bressanelli G. & Saccani N., 2021). Our study examined 600 manufacturing companies across various sectors and assessed their Circularity scores. A total of 380 companies spanning the textile, electronics, oil and gas, and general manufacturing sectors were selected for analysis.

This report provides an in-depth analysis of the C-readiness score of 93 valid responses from textile companies in our sample. Evaluating the circularity scores of textile Industry provides valuable insights into their readiness for adopting CE practices. Further, we conducted a detailed descriptive analysis based on the collected data to understand the scores in the different dimensions.

To gain a deeper understanding of the barriers to adopting CE practices in the Indian textile industry, we conducted a narrative literature review to analyse the challenges faced. This analysis helped in synthesize the current scenario of the Indian textile industry in transitioning towards a CE. This assessment provides a roadmap and guidelines for the industry to effectively adopt CE practices in the Indian context. From an interdisciplinary perspective, we discussed the digital technologies that can help to overcome the identified barriers and enhance the circularity of the textile industry. By outlining digital technology requirements, we aim to provide a roadmap for the industry to leverage digital solutions and overcome the barriers to adopting CE practices.

The remainder of this paper is structured as follows: In the next section, we describe our methodology. Then, we present a detailed descriptive statistical analysis and the results of the sample, followed by a comprehensive discussion of the barriers for the Indian textile industry that hinder the adoption of the CE. Finally, we discuss a digital pathway framework for overcoming these barriers and conclude the article.

## Methodology:

To ensure methodological rigor, we employed a two-stage approach. In the first stage, we conducted a survey using the C-readiness tool, which allowed us to analyse six specific dimensions for understanding the adoption of the CE. This tool provides a comprehensive framework to assess an organization's readiness in transitioning towards circularity. The six dimensions examined were material and product circularity, business model circularity, organizational circularity, supply chain circularity, production circularity, and end of life (Bressanelli G. & Saccani N., 2021). By utilizing this tool, we gained a deeper understanding of the current state of CE adoption within the Indian textile industry, identifying both strengths and barriers. Then, we aim to identify and describe the barriers to the adoption of the CE in the Indian textile industry using a narrative literature review methodology. Unlike a systematic review, this approach does not follow a strict protocol but instead allows for the identification of the key studies that explore the issue of interest (Chaney, 2021). The aim is to identify and describe the barriers to the adoption of the CE in the textile industry using a narrative literature review methodology (Ardolino et al., 2022). This methodology has already been employed by other scholars discussing sustainability and CE issues (Chofreh et al., 2019; Upadhyay et al., 2021). It provides a comprehensive understanding of the barriers to widespread adoption of CE practices in the Indian textile industry, without the constraints of a strict systematic review. To identify the articles to be analysed, we searched the Scopus and Web of Science databases using keywords related to CE, and barriers. The search string is used as (‘Circular economy’ OR ‘sustainability’) and related to barriers (‘Barrier*’ OR ‘Challenges’ OR ‘Obstacles’). The literature focused on the Indian context was selectively analysed to provide a more comprehensive understanding of the barriers to widespread adoption of CE practices in the Indian textile industry. The research methodology is represented in Figure 1.

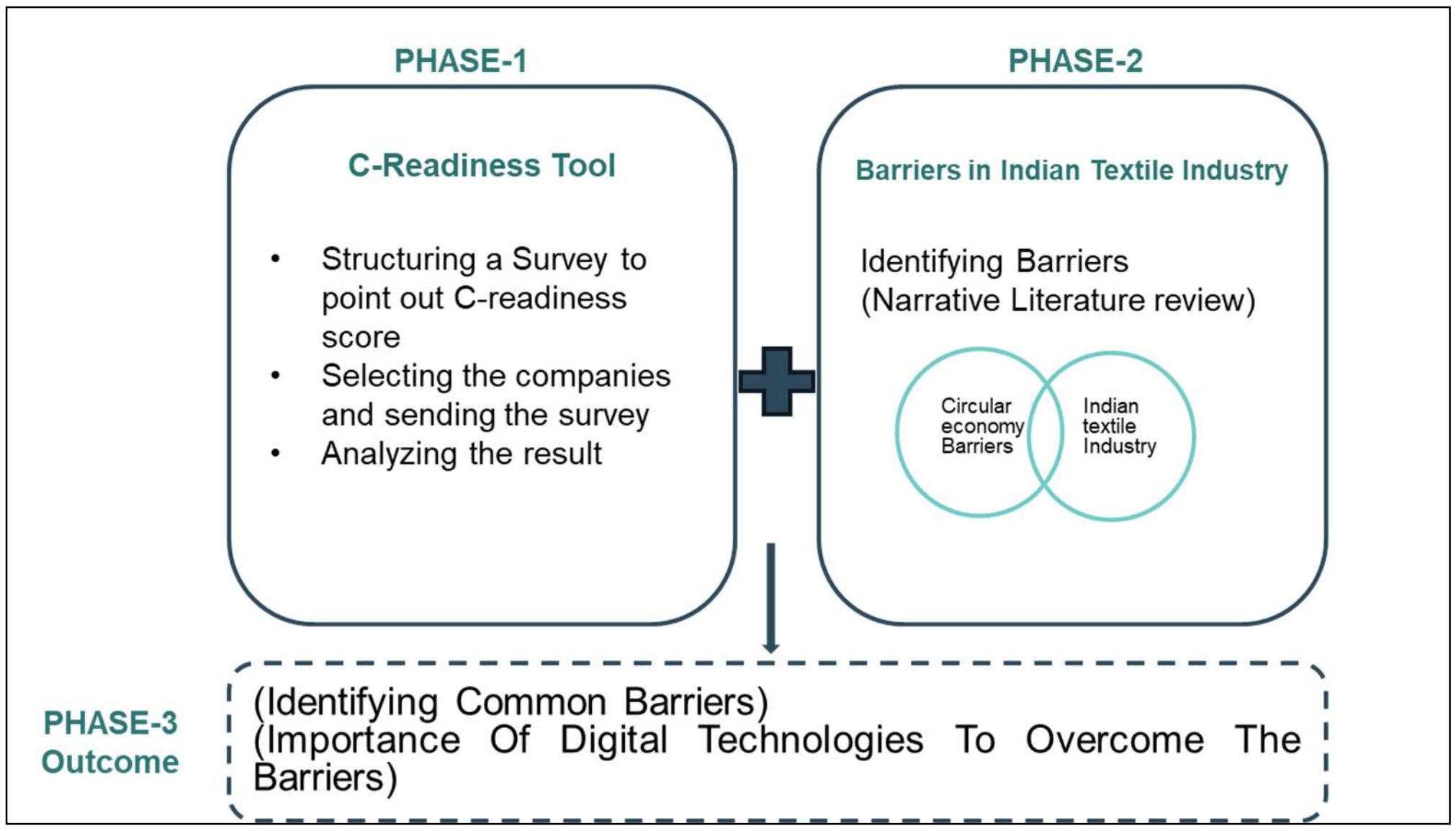

*Figure 1 Research methodology*

**C-Readiness Tool:**

Initiating a systematic transition towards a CE requires companies to first assess their existing practices. However, many companies struggle to start this transition due to a lack of structured assessments to guide their significant transition processes (Sohal et al., 2022). To fill this gap, the C-Readiness tool provides a systematic and structured framework to help companies assess their preparedness and readiness for this transition. The C-Readiness tool allows companies to thoroughly evaluate their existing practices against established CE benchmarks. This assessment encourages collaboration not only among different departments within the organization, but also with external partners. This promotes a common understanding of CE goals across the organization. Additionally, the tool enables companies to assess their performance against industry standards and peers, which promotes ongoing improvement and innovation in circular practices. By employing this comprehensive framework, organizations can make informed decisions and develop tailored strategies to effectively transition towards a CE model.

The C-Readiness tool systematically evaluates the preparedness of manufacturing industries for adopting a CE by assessing six key dimensions that determine their circularity readiness (Bressanelli G. & Saccani N., 2021). The tool also supports the integration of Life Cycle Assessment data to provide a holistic view of circularity, enabling organizations to understand the linkages between CE practices and decarbonization. The six major dimensions assessed by the tool facilitate a thorough evaluation of a company's current state and readiness level as shown in Figure 2.

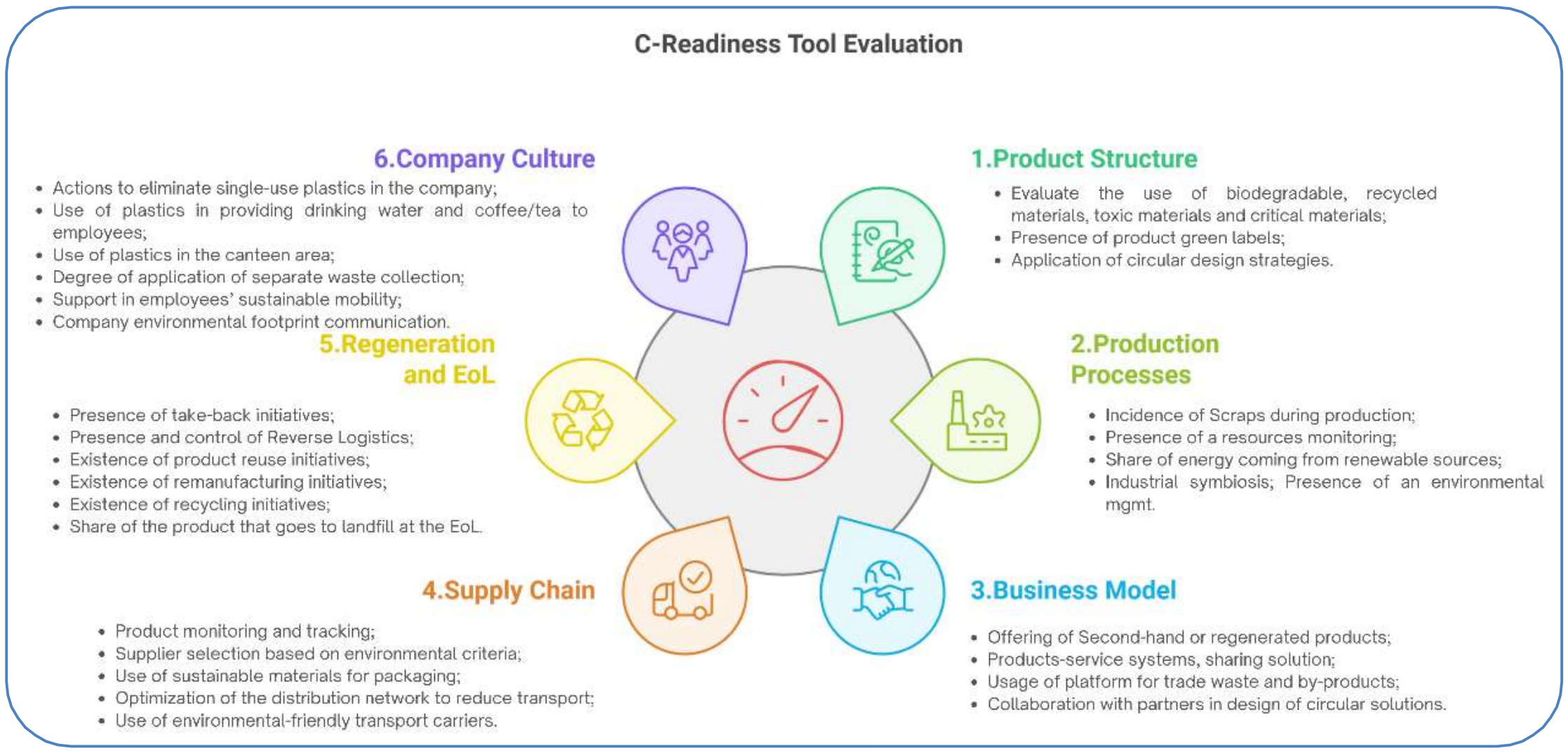

*Figure 2 C-Readiness evaluation elements*

After completing the assessment, the tool typically provides a readiness score that precisely reflects the organization's current state. Based on these comprehensive assessment findings, the tool then provides a tailored action plan that outlines the specific steps needed to further enhance the company's circularity. This action plan includes detailed guidance on resource allocation and the implementation of targeted initiatives for improvement. For instance, the tool may recommend that companies prioritize investments in areas with high environmental impact but low C-Readiness scores, while potentially neglecting areas with low environmental impact and high C-Readiness scores.

**Result:**

India is a major global textile and garment producer, accounting for 4% of worldwide trade in these sectors. The country is projected to reach $250 billion in textile production and $100 billion in exports by 2030 (Ankita Sharma, 2020; Ncaer, 2009). India's textile industry is anchored by regional clusters in the northern, southern, and western belts that blend traditional craftsmanship with modern production to meet global demands (CII, 2023).

India's vast industrial sector makes it challenging to include all companies in the survey. For this study, we have focused on the western industrial belt. The application of the C-Readiness tool enabled a detailed assessment of the overall circularity score and performance across individual areas for each of the 93 companies in the sample. The results show that 96% of the companies belong to SMEs sector, while only 3% are from the large industries as shown in Figure 3.

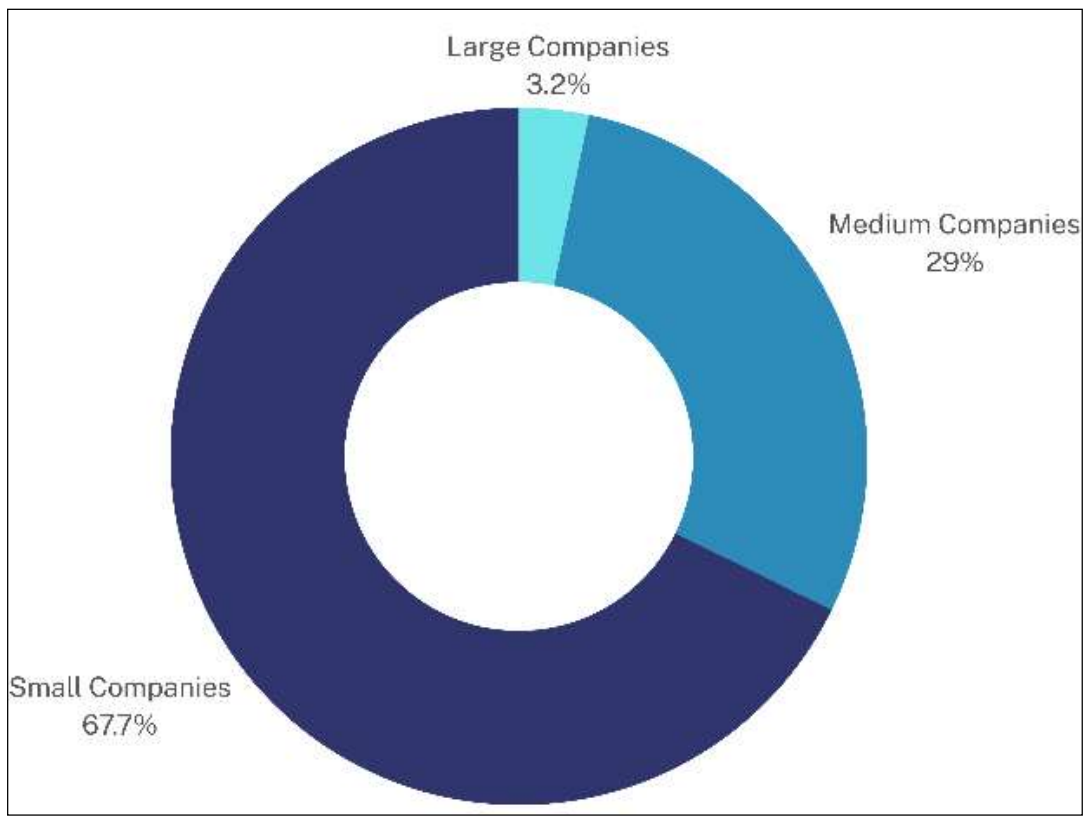

*Figure 3 Distribution of companies that participated in the survey*

The survey data depicted in Figure 4 represents the job roles of the respondents. The majority were owners or senior-level managers, suggesting that the data reflects the perspectives and insights of the leadership responsible for key strategic and operational decisions within these organizations. To clarify the responses and analysis, we conducted an hour-long discussion with the participants, who were primarily owners or top managers of the companies. This allowed us to better understand their perspective on the circular economy.

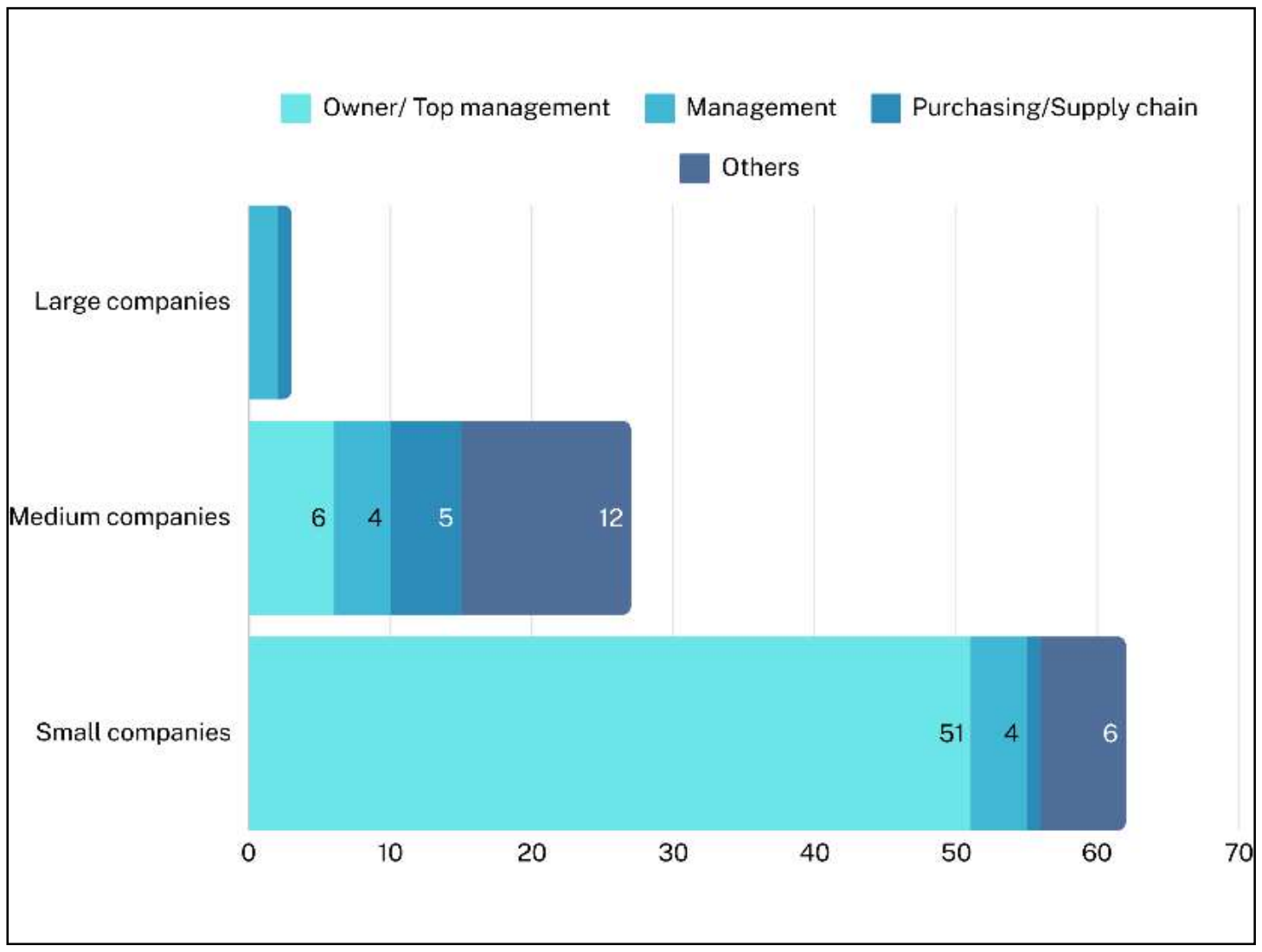

*Figure 4 Distribution of job roles among survey participants*

The average circularity score of the analysed sample is 43%, highlighting that Indian textile companies exhibit a relatively low level of 'readiness' for the CE paradigm, and significant potential for improvement in aligning with CE principles as shown in Figure 5.

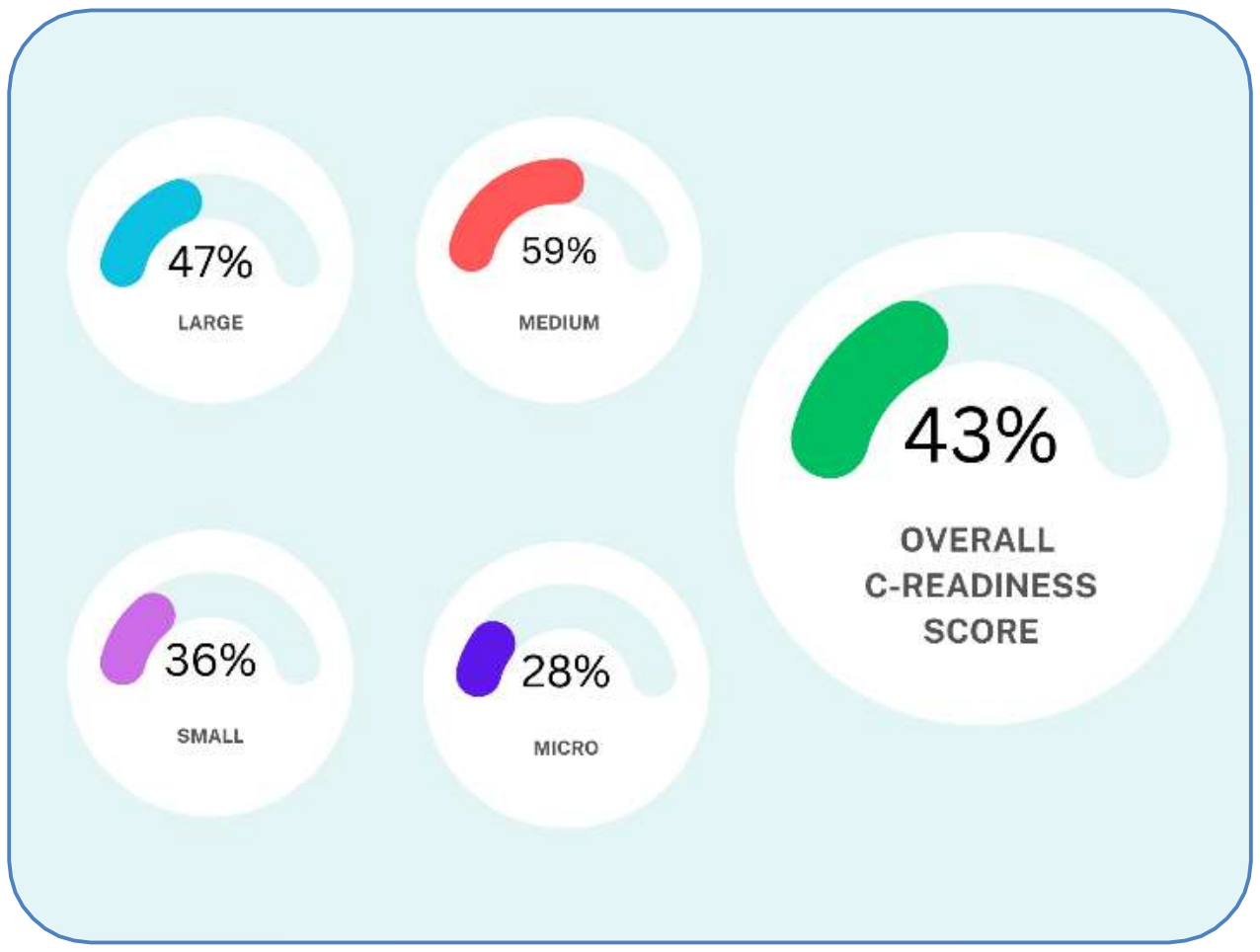

*Figure 5 The overall C-Readiness Score*

The analysis shows that 39% of the companies have circularity scores below 25, indicating that the Indian textile production ecosystem still has significant progress to make in fully adopting and applying CE principles and practices.

The analysis further examines the six key areas of CE implementation in Figure 6. The sub-dimensions of the six dimensions are presented in Appendix A1-A6. The analysis indicates that companies show the highest adoption of circularity principles in the supply chain domain, with a score of 49%. This suggests that companies in India have implemented innovative practices to support a closed-loop supply chain for textile recycling. Most of the industry in India has a semi-organized process for recycling cutting waste, which presents both challenges and opportunities (Dwivedi et al., 2023; Ponnambalam et al., 2023). For instance, the Re-Start Alliance initiative aims to commercialize advanced textile recycling technologies and establish a sustainable supply chain for textile waste feedstock. This allows for the effective collection, sorting, and processing of post-consumer textile waste, enabling the reintroduction of recycled materials back into the manufacturing process (Nickerson, 2024).

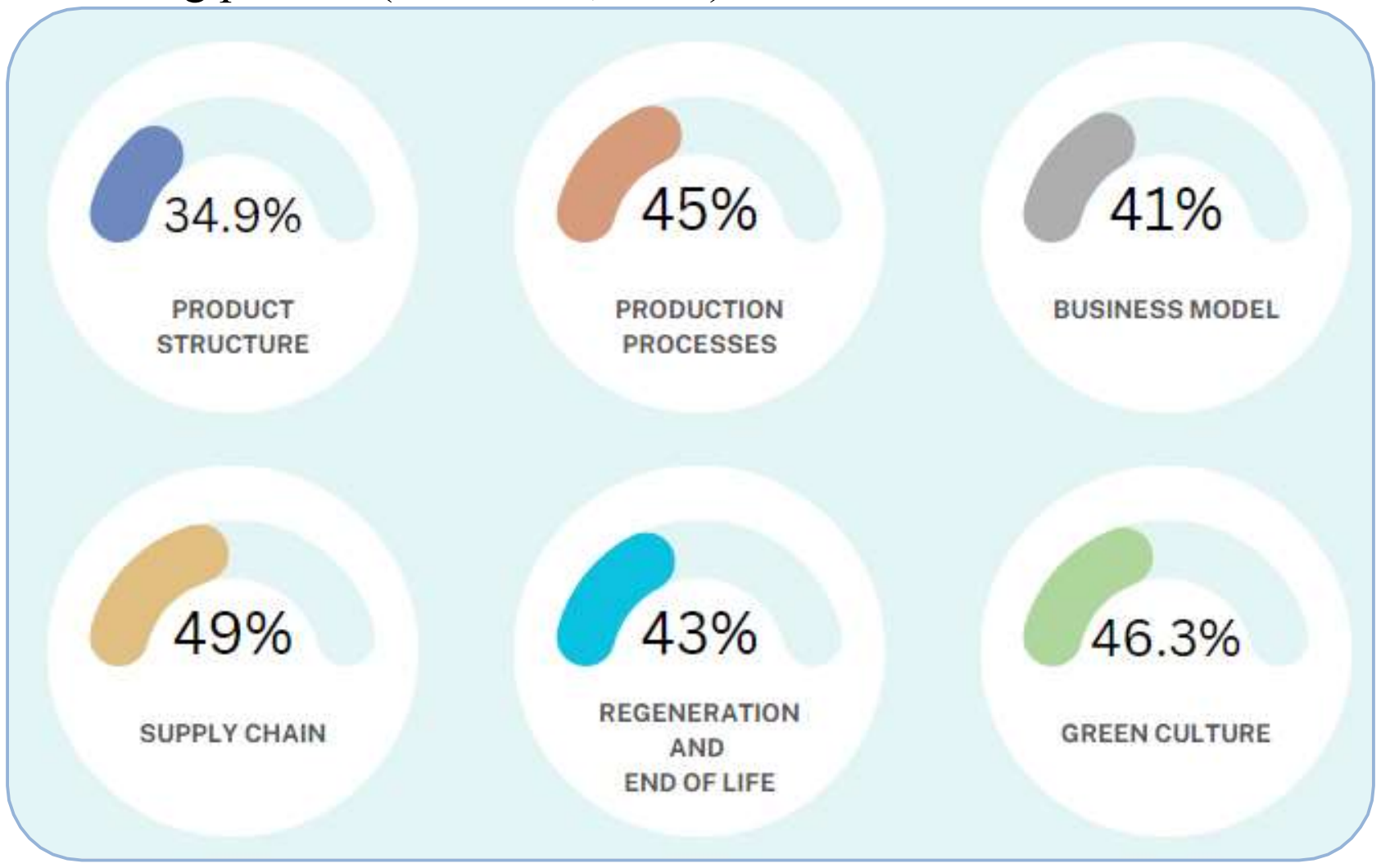

*Figure 6 C-readiness Score Based on Six dimensions.*

Regarding product structure, the average circularity score is 34.9%. It was found that 40% of companies do not use biodegradable, recycled, or recyclable materials. One challenge is highlighted by Dwivedi et al., (2023) that the production of coloured selvage waste during the weaving of dyed yarns reduces the recyclability of the material. Additionally, 52% of companies utilize critical materials in their products. This statistic highlights a critical gap in the adoption of sustainable materials, indicating a need for greater awareness and education about the benefits of circular materials and their long-term cost savings. Meanwhile, 40% of companies lack environmental product certifications, although they intend to obtain them. The absence of certifications may undermine consumer trust and limit access to markets increasingly driven by sustainability criteria. Dwivedi et al., (2023) emphasized the need for companies to proactively obtain certifications such as OEKO-TEX, GOTS, BSCI, SLCP, GRS, Bluesign, and ZDHC to promote circularity (Dwivedi et al., 2023). However, it is observed that small- to large-scale textile industries predominantly processed based on vendor specifications and product's economic value, without considering environmental regulations, and standards. On the issue of toxic materials, a significant number of companies (52%) recognize the need to eliminate the use of toxic materials but continue to use them. Recognizing the need to eliminate toxic materials is crucial for improving product safety and reducing environmental harm, but companies must develop and implement actionable strategies to effectively phase out these materials.

As far as business model concerns the average score is the second lowest at 41%. The data reveals that only 9% of companies are aware of platforms for buying and selling production scrap and rejects. This indicates a significant lack of awareness about practical tools and platforms that can facilitate such collaboration. However, there is scope for innovative tools, such as scrap trading platforms, to help companies effectively manage their waste. Additionally, a large number of companies (38%) recognize the need to collaborate with suppliers and/or partners across the value chain to co-design products and processes with a circular approach. By engaging in co-design efforts, sharing information transparently, fostering joint innovation, managing risks collaboratively, and aligning on shared goals, companies can significantly enhance their sustainability initiatives and contribute to a more CE (Nudurupati et al., 2022). When it comes to end-of-life generation, only 7% of companies encourage consumers to support the recovery of end-of-life products, components, and materials through a structured process. This observation indicates that companies lack organized initiatives to recover end-of-life (EoL) products and components, revealing systemic challenges within the industry regarding EoL product management. Therefore, companies should invest in developing structured recovery programs that clearly outline processes for collecting and processing EoL products (Molla, 2022). To increase consumer participation in EoL recovery, companies should implement awareness campaigns that highlight the benefits of returning products at the end of their life cycle. Utilizing technology can also streamline recovery processes and improve consumer engagement. For example, utilizing blockchain to streamline product returns or track recycling efforts can improve the user experience. Additionally, offering instant loyalty points for returning end-of-life products can increase the consumer participations.

In the production domain, the Indian textile sector exhibits significant gaps in monitoring its resource consumption. Only a small fraction, merely 3%, of companies continuously monitor and periodically take action on their system consumption (energy, water, and air) suggesting a lack of robust monitoring systems (Saccani et al., 2023). While the companies recognize the significance of renewable energy sources, a staggering 53% have yet to adopt them, hindering their progress towards sustainability. Another concerning finding is that 24% of the companies are unaware of the concept of industrial symbiosis, a crucial strategy within the CE framework. Industrial symbiosis involves the collaborative exchange of materials, energy, and services between companies, allowing one company's waste to become another's resource. Companies unaware of industrial symbiosis are likely missing out on opportunities for cost savings, innovation, and enhanced sustainability. This lack of awareness underscores the need for greater education and awareness-building initiatives to enable the adoption of CE principles across the textile industry.

**Barriers:**

To understand the low CE readiness in the Indian textile sector, a narrative literature review was conducted. Since very few studies specifically examined CE adoption barriers in the textile industry, the review focused more broadly on barriers in Indian manufacturing. Key barriers relevant to the textile sector were then identified by comparing CE-readiness tool scores. So, the narrative literature review identifies macro, meso, and micro-level barriers that impede the adoption of the CE in Indian manufacturing. Prior research has identified common barriers to implementing the CE in supply chains, including barriers related to sustainability benefits such as social, economic, and environmental factors, or cultural and regulatory factors (Das et al., 2024). Nudurupati et al., (2022) study highlights the barriers faced by Indian SMEs in adopting CE practices; Key challenges include limited access to technology and expertise, which hinder the effective implementation of CE strategies. The study also underscores the role of limited consumer awareness as a significant barrier to driving demand for CE practices among SMEs. Rajput & Singh, (2021) identified additional barriers specific to Industry 4.0 and CE integration in India, such as process digitalization challenges, semantic interoperability issues, and insufficient technical knowledge. Bag et al., (2022), in his case study, emphasized 35 barriers to digital manufacturing and CE adoption in India, including limited digital infrastructure, digital literacy gaps, and concerns over data security. Leadership deficiencies, such as unclear digital strategies, and workforce-related challenges, such as inadequate training, further complicate the adoption of CE practices. Yadav et al., (2020) categorized CE barriers in Sustainable Supply Chain Management into five groups, pinpointing economic, organizational, and supplier-related challenges. Chhimwal et al., (2022) emphasized the interconnection between barriers, such as regulatory limitations affecting market demand, which in turn discourages technological investments. For instance, unclear legislative frameworks can create a cascade of challenges impacting technological readiness. Dutta et al., (2021) focused on reverse logistics, identified 23 barriers and stressed the need for a structured approach to address them. Badhotiya et al., (2022) studied the CE barrier into three perspectives. The challenges faced by industries in transitioning to a circular model are complex and cannot be fully captured by a simple three-category. Hence, the multifaceted nature of the CE barriers often leads to overlap and interplay, making them difficult to categorize. Addressing these barriers requires comprehensive frameworks that consider a broader range of factors beyond just social, economic, and environmental considerations. The literature collectively suggests that CE barriers vary significantly across sectors and regions, shaped by unique economic, cultural, and regulatory contexts. Cultural perceptions and traditional business practices also significantly impact CE adoption, especially in developing regions. Resistance to transitioning from linear to circular models is often rooted in cultural norms and established practices. In similar note, cultural (stakeholder) perceptions significantly influence the adoption of circular economy practices, as challenges such as limited understanding of CE concepts, perceived risks in transitioning from linear to circular business models, and insufficient collaboration among supply chain partners particularly for SMEs in India (Mangla et al., 2018; Sharma et al., 2024). A critical barrier frequently identified in studies is the deficiency of CE knowledge within organizations and among consumers (Dutta et al., 2021; Sharma et al., 2023). Although financial constraints are frequently cited as a key challenge, these studies explore more specific factors within the financial domain, such as the costs of raw materials or the availability of recycling infrastructure, which may be highlighted in a more context-dependent manner (Sharma et al., 2024; Zaidi & Chandra, 2024).

Furthermore, sectoral and geographical variations introduce unique barriers (Gedam et al., 2021). For instance, metal industries struggle with financial and infrastructure constraints (Badhotiya et al., 2022), while the mining sector lacks adequate regulatory measures for efficient waste management (Badhotiya et al., 2022). In contrast, the plastic and automotive industries encounter barriers related to technological limitations, low consumer awareness, and insufficient government support (Bag et al., 2022; Manoharan et al., 2022). Market-related challenges persist as well, including fears of cannibalizing existing product lines, potential loss of proprietary knowledge during recycling processes, and concerns about brand reputation due to perceptions of lower-quality recycled materials (Fiksel et al., 2021). Technological barriers, such as limited access to advanced recycling systems and weak waste management infrastructure, compound these issues (Parida et al., 2023). Additionally, weak regulations and poor enforcement of existing policies further hinder CE adoption, particularly in India. The lack of clarity regarding responsibilities within supply chains, combined with inadequate governmental support for sustainable practices, creates an uncertain business environment (G. Yadav, Luthra, Kumar, et al., 2020). Despite these insights, existing studies often fail to address the dynamic nature of these barriers, which can evolve over time, thereby necessitating continuous assessment to identify the new barriers.

Based on the comprehensive review of these literature, the key barriers to the adoption of CE principles in Indian manufacturing industries can be categorized into 9 main categories, which are further divided into subcategories as presented in the Appendix B1.

Although several studies discuss CE adoption barriers broadly, a comprehensive understanding of the unique challenges faced by Indian textile industries remains limited. Previous research, such as Zaidi & Chandra, (2024) study on India's textile manufacturing sector, has highlighted some of these challenges, which align closely with the findings of Saccani et al., (2023). To deepen our understanding of these barriers, we adopted and revised Saccani et al., (2023) categorization framework. The study then explores the complex interconnections among CE adoption barriers for Indian textile industries, comparing these findings with those from the CE-readiness assessment tool. Through cross-impact matrix analysis, we identified critical barriers, their interdependencies, and potential cascading effects. Table 1 presents the Barriers Interdependency Framework, which is developed based on a comparison between the narrative literature review and an assessment of c-readiness. The framework highlights the significance of challenges across the six key dimensions of the c-readiness evaluation tool.

This analysis highlights several critical barriers to the adoption of CE practices in textile industry. In the Indian context, financial constraints and limited technical resources and infrastructure emerge as prominent challenges, often influenced by local economic and infrastructural conditions. Financial viability is a significant barrier, as many organizations face high investment requirements and budgetary limitations to adopt CE practices (Badhotiya et al., 2022; Nudurupati et al., 2022). Furthermore, the decoupling of costs and perception of low returns and revenues delays acceptance of new business models like servitization (Saccani et al., 2023). Lack of infrastructure, process management and technological challenges encompass difficulties in recycling post-consumer textiles while maintaining fibber quality, poor labour standards, inadequate tools for assessing return flows, and risks in outsourcing remanufacturing processes (Fiksel et al., 2021; Ponnambalam et al., 2023). Additionally, Limited expertise in recycling within unrecognized industries has led to pre- and post-consumer textiles being classified as "special waste" instead of by-products or secondary raw materials (Saccani et al., 2023). This misclassification restricts their reuse potential and contributes to most fabrics being downcycled into lower-value products such as carpets or towels, undermining efforts toward sustainable resource use.

Policy and regulatory barriers represent a critical challenge in advancing the CE (V. S. Yadav & Majumdar, 2024). The absence of a national quality benchmarking system prevents effective evaluation and comparison of circular practices across different sectors, leading to inconsistencies in implementation. Regulatory pressures further complicate this landscape, as existing policies often fail to support or incentivize circular initiatives (Ponnambalam et al., 2023). Misalignment between current regulations and the principle of the CE creates significant obstacles. A cohesive framework is essential for coordinating efforts among various stakeholders, including government agencies, businesses, and consumers (Mhatre et al., 2023). Without such coordination, initiatives may become fragmented and less effective. The paper further explores barriers such as the challenges of adopting renewable energy in the textile industry, as well as the lack of coordination and information sharing among supply chain partners (Zaidi & Chandra, 2024). These issues are particularly due to insufficient financial resources and limitations in infrastructure development. Although data security and privacy are not currently viewed as significant obstacles, their importance is likely to grow with the increasing integration of smart textile products (Khan et al., 2022; Mangla et al., 2018).

The rapid change of fashion trends further complicates the integration of durable products with advancing technologies. Finally, consumer behaviour poses another barrier, as many consumers lack confidence in the quality and reliability of products marketed as sustainable or circular. This scepticism can limit demand for such products, making it challenging for companies to justify investments in circular practices (Yaduvanshi et al., 2016).

# Table for barriers

*Table 1 Integrated Framework*

| Categories | Barriers | Product structure | Production processes | Business model | Supply chain | Regeneration & End of life | Corporate & Good culture | Total |
|---|---|---|---|---|---|---|---|---|
| Financial or economic Barriers | Lack of high capital investment (B1) | | X | | | X | X | VERY HIGH |
| | Lack of financial support for CE adoption (B2) | | X | | | X | | MEDIUM |
| | Lack of competitive virgin material pricing (B3) | X | | | | | | HIGH |
| | Perceived low returns and financial risks (B4) | | | | | X | X | MEDIUM |
| | Lack of access to affordable green technology (B5) | | X | | | | | LOW |
| Operational Barriers | Lack of product design strategies for reuse and remanufacturing(B6) | X | | X | X | X | | VERY HIGH |
| | Product complexity (B7) | X | X | | X | | | HIGH |
| | Lack of design strategies for energy consumption reduction (B8) | | | | | | | LOW |
| Organisational Barriers | Lack of effective employee training and engagement for circular economy (B9) | | | | X | | X | LOW |
| | Lack of cultural alignment with circular economy principles (B10) | | | | | X | | LOW |
| | Lack of top management motivation and resistance to change (B11) | | | | X | X | X | MEDIUM |
| | Lack of industrial support and industrial symbiosis (B12) | | X | X | | X | | HIGH |
| Market Barriers | Lack of a business plan for financial returns (B13) | | | X | | | | LOW |
| | Lack of strategies to address product cannibalization (B14) | | | X | | X | | LOW |
| | Lack of access to know-how and expertise (B15) | | | X | | X | | HIGH |
| | Lack of market availability for reuse products (B16) | | | X | | X | | LOW |
| Technological Barriers | Lack of advanced technology and innovation (B17) | X | | | X | X | | HIGH |
| | Lack of data privacy and security measures (B18) | | | | | | | |
| | Lack of traceability in supply chains (B19) | | | | X | | | LOW |
| | Lack of infrastructure and resource capacity (B20) | | X | X | X | X | | VERY HIGH |
| Environmental Barriers | Lack of awareness regarding environmental impacts (B21) | X | X | | | X | X | MEDIUM |
| | Unorganized recycling industry (B21) | | | | X | X | X | HIGH |
| | Lack of an appropriate system for reverse logistics (B22) | | | X | X | X | | MEDIUM |
| Regulatory Barriers | Lack of taxation and incentives aligned with circular economy goals (B23) | X | | X | | | X | HIGH |
| | Lack of policies, laws, and standard systems for CE (B24) | | | X | | | | LOW |
| | Lack of enforcement ability for existing legislation (B25) | X | | | | X | | LOW |
| | Lack of alignment between measures, metrics, and CE indicators (B26) | X | | X | X | | X | HIGH |
| Supply chain (SC) Barriers | Lack of effective communication with suppliers (B27) | | | X | | X | | LOW |
| | Lack of coordination and information sharing among stakeholders (B28) | | | X | X | X | X | HIGH |
| | Lack of effective supplier selection strategies (B29) | | | | X | | | LOW |
| | Lack of supply chain designs for reuse and remanufacturing (B30) | | | X | X | X | | MEDIUM |
| Cultural Barriers/Social Barriers /User Behaviour challenges | Lack of awareness and knowledge among consumers and suppliers (B31) | | | X | X | X | X | VERY HIGH |
| | Lack of positive consumer perception towards circular economy practices (B31) | X | | X | | X | | VERY HIGH |
| | Lack of strategies to address the loss of ownership value (B32) | | | X | X | | | LOW |
| | Lack of encouragement for design for disassembly and recycling (B33) | X | | X | | X | | MEDIUM |
| | Lack of measures to mitigate circular rebound risks (B34) | | | | | | | |

**Digital Technology To support The Barriers:**

Industry 4.0 technologies offer transformative solutions that address key barriers to implementing CE practices in the textile sector (Kanupriya, 2021). For example, the lack of traceability in India's textile industry poses a significant challenge in effectively managing the more than 1 million tons of textile waste discarded annually (Dwivedi et al., 2023). However, the implementation of advance traceability systems leveraging Industry 4.0 technologies can enable the tracking of textile materials from origin to end of life, thereby facilitating more efficient recovery, reuse, and recycling effort (Saccani et al., 2023). This, in turn, can improve recycling processes and enhance the overall management of textile waste.

However, organizations often face difficulties in integrating digital technology solutions with their existing systems due to the complexity of integration, which can lead to interoperability issues between different technologies and systems (Bressanelli et al., 2022). Additionally, many of these technologies heavily depend on data for effective data management for successful implementation, such as generating data or needing data to derive actionable insights(Bag et al., 2020). Furthermore, ensuring the quality and accuracy of data collected from various sources can create challenges in implementation. In addition to it, different technologies require varying levels of investment and resources which While the companies are unwilling to invest in new technologies due to uncertainty about the potential Return of investment creating a negative perception (Ingemarsdotter et al., 2021). Hence, understanding the mature, niche, and emerging technologies relevant to the CE is vital because understanding the level of maturity helps mitigate risk associated with technology adoption. For example, Mature technologies generally have lower risks associated with implementation due to established frameworks and case studies. By understanding the maturity of these technologies, companies can make informed decisions about where to allocate their budgets effectively (Neri et al., 2023). Because Mature technologies might offer quicker returns on investment, while emerging technologies could represent long-term strategic investments. A comprehensive understanding of the technological landscape enables organizations to develop strategic roadmaps for their transition towards CE models. Furthermore, knowing which technologies are mature versus emerging helps organizations establish metrics for evaluating their impact on sustainability goals. By understanding the technologies required to accelerate the CE's adoption, organizations can fully leverage initiatives like Digital India and digital strategies to achieve their goals.

In the context of textile manufacturing, the adoption framework discussed in Figure 7 discussed the technologies used in the textile industry as mature, niche, or emerging. Mature technologies for example, by integrating IoT and data analytics capabilities, companies can monitor the lifecycle of textiles from production to end-of-life, ensuring that all stakeholders have access to reliable data regarding material origins and transformations. These technologies enhance traceability, enable mass customization, manage product complexity, foster consumer loyalty, optimize return flows, facilitate partnerships, and improve coordination among stakeholders (Gupta et al., 2021; Jamwal et al., 2022). Utilizing data analytics and artificial intelligence can also streamline production processes. These technologies can analyse energy consumption data to improve production processes and understand consumer trends and preferences in real-time, enabling companies to adapt quickly to changes in demand or fashion trends while maintaining a focus on sustainability (Liang et al., 2018). Additionally, advanced data analytics can enable better forecasting and inventory management, allowing companies to analyse patterns in returns and adjust their logistics, accordingly, reducing uncertainty associated with post-consumer textile returns.

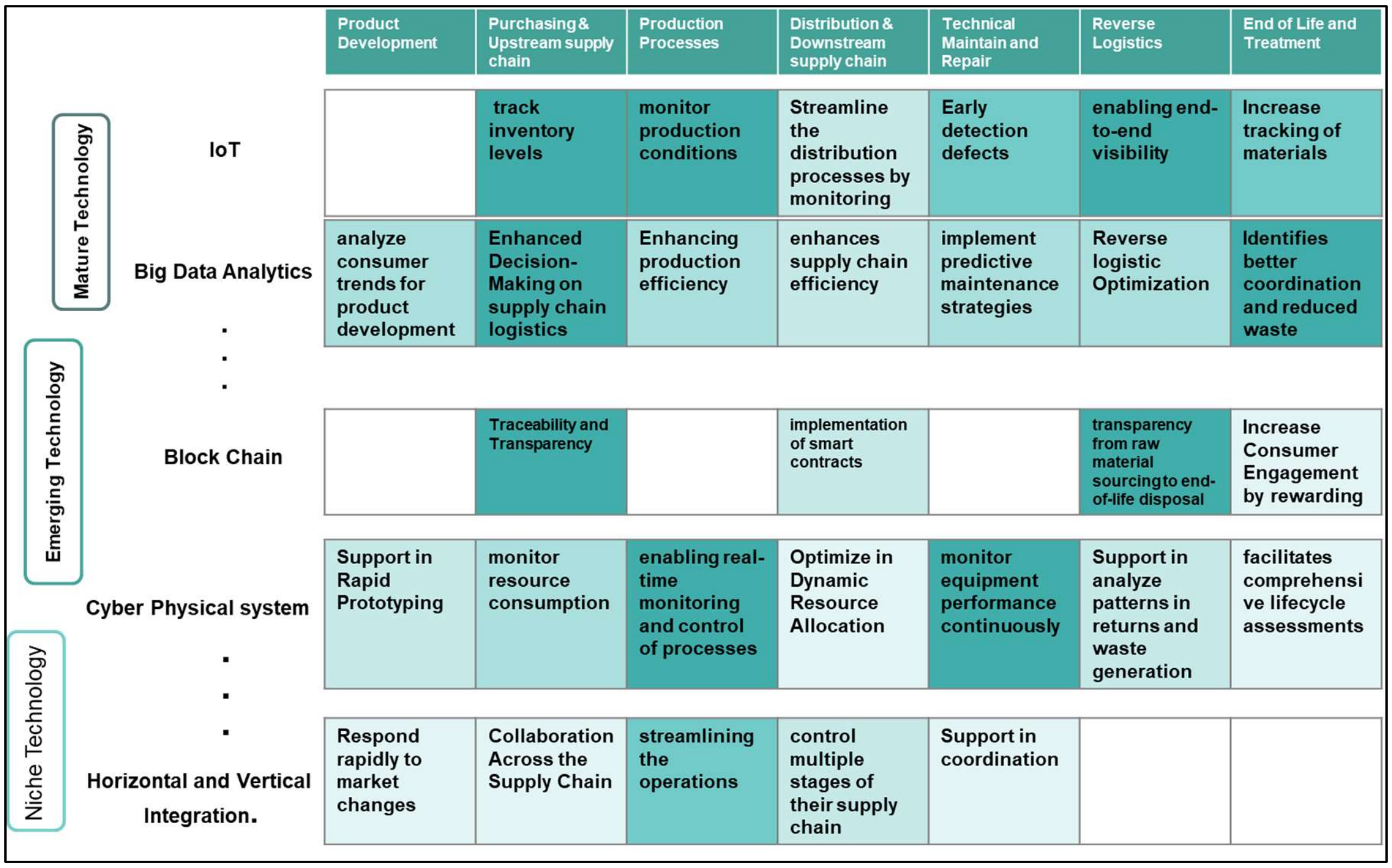

*Figure 7 Digital Roadmap for Enabling Circular Economy in the Indian Textile Industry*

Emerging technologies, on the other hand, represent innovative approaches that have the potential to disrupt and transform the textile industry in the future (V. S. Yadav & Majumdar, 2024). Advanced manufacturing technologies, such as robotics, cyber physical system, additive manufacturing and block chain can reduce waste in the textile industry. For example, robotics and cyber-physical systems can automate the textile cutting process, thereby enhancing efficiency compared to manual cutting methods. Through cyber-physical systems, the digitalization of logistics and inventory can be supported by collaborative transportation and optimized transportation routes, leading to maximum reutilization of packaging material and minimization of waste. The network-based, streamlined communication channels of these technologies enhance collaboration and reduce delays by facilitating integration between different organizations or firms, creating a more cohesive supply chain. Additionally, 3D printing technology enables the production of customized textile products on demand, reducing waste from overproduction. This technology promotes self-reliance in manufacturing as it becomes more affordable, providing greater access to customized 3D textile products such as buttons and other accessories (Hettiarachchi et al., 2022; V. S. Yadav & Majumdar, 2024). Further, Blockchain technology enhances the CE in textile industries by improving data security through decentralization and immutable records, reducing the risk of data breaches and unauthorized access. It provides a common platform for data sharing, fostering collaboration among stakeholders through shared data access and supply chain traceability. Additionally, blockchain enables loyalty token programs that reward consumers and businesses for participating in CE practices, such as recycling, returning products, or using sustainable materials (Esmaeilian et al., 2020; Prajapati et al., 2022).

Similarly, niche technologies can be more complex but may address specific needs or sectors within the CE. SMEs can benefit from customized niche solutions tailored to their unique operational contexts. For instance, horizontal and vertical integration, augmented reality, or virtual reality can improve cross-industry coordination and communication. Horizontal and vertical integration can connect different levels of the supply chain, from raw material suppliers to end consumers. These technologies can also facilitate seamless collaboration between manufacturers, suppliers, retailers, and consumers, ensuring that all parties are aligned in their efforts toward circularity (Schöggl et al., 2024). By strategically adopting these technologies based on their maturity levels, textile manufacturers can drive the transition towards a CE.

## Conclusion:

The successful implementation of CE principles within India's textile industry requires a interdisciplinary approach. This approach should leverage the findings provided by the C-Readiness Tool, which access the critical areas of the textile industry, identifying strengths, and providing recommendations based on current practices. However, the readiness of the Indian textile industry is relatively low, around 43%. This indicates that there is still a long way to go to fully adopt the CE in Indian textile industries. Nevertheless, the C-Readiness Tool provides organizations with a comprehensive assessment and a customized action plan to facilitate their transition towards a CE. While existing literature has extensively examined barriers to adopting Circular Economy principles in Indian manufacturing industries, there remains a gap in systematically analysing and categorizing these barriers within the context of the Indian textile sector. By identifying and analysing 34 barriers, this review offers a detailed taxonomy that provides critical understanding for policymakers, industry stakeholders, and academic researchers on the actions necessary to support the transition towards a more sustainable and circular textile ecosystem in India. Furthermore, this paper emphasizes the integration of digital technologies as essential tools to overcome existing challenges. It underscores the importance of addressing sector-specific needs to enable strategic planning and tailored solutions.

Additionally, it highlights the importance of mature, niche, and emerging technologies relevant to the CE, based on industry-specific requirements. This paper is one of the first to discuss digital solutions based on technology maturity levels, enabling the industry to gain a comprehensive understanding of the technology landscape. These aspects help companies prepare their workforce for changes, ensuring employees are upgrade the necessary skills and knowledge to adapt. Moreover, understanding the technology landscape allows companies to engage effectively with partners, suppliers, and customers by aligning on technology adoption strategies that support circularity goals. This paper serves as a call to action for all stakeholders in the Indian textile sector to actively embrace CE principles. Consequently, a comprehensive examination of the interconnected roles and complementary nature of various digital technologies in the CE is necessary to gain broad understanding, supporting to more practical strategies for resource management, waste reduction, and sustainable production.

## Managerial Implication

The study offers managerial implications for Indian textile managers and policymakers, providing insights into the complex and dynamic efforts of the Indian textile industry toward adopting CE principles at a micro level. The study highlights a critical finding - the pressing need to increase awareness of CE practices, challenges, and associated benefits. To address this, governments and other key stakeholders should undertake targeted initiatives such as workshops, seminars, and informal engagement sessions. Furthermore, governments must take a more active role, providing financial incentives like micro-loans and tax cuts, while also accelerating the establishment of eco-industrial parks and implementing supportive CE policies. In addition to this, increasing consumer awareness is crucial for the acceptance of eco-friendly and circular textile products. Targeted educational campaigns can help shift consumer preferences, fostering sustained demand for sustainable textiles and supporting the industry's transition to circularity. The study also highlights the transformative potential of collaborations between large enterprises and SMEs. By financially supporting SMEs and aligning their sourcing strategies with sustainability goals, large businesses can ensure a steady supply of raw materials derived from SME by-products. This partnership not only benefits SMEs but also enhances resource efficiency for large corporations. Finally, Indian textile SMEs should take a more active role in the CE by collecting and segregating used materials, thereby reducing their dependence on virgin resources. This proactive approach would significantly advance the industry's transition to a circular textile ecosystem.

# Appendix

Figure A1: Categorization of Responses Based on Product Structure

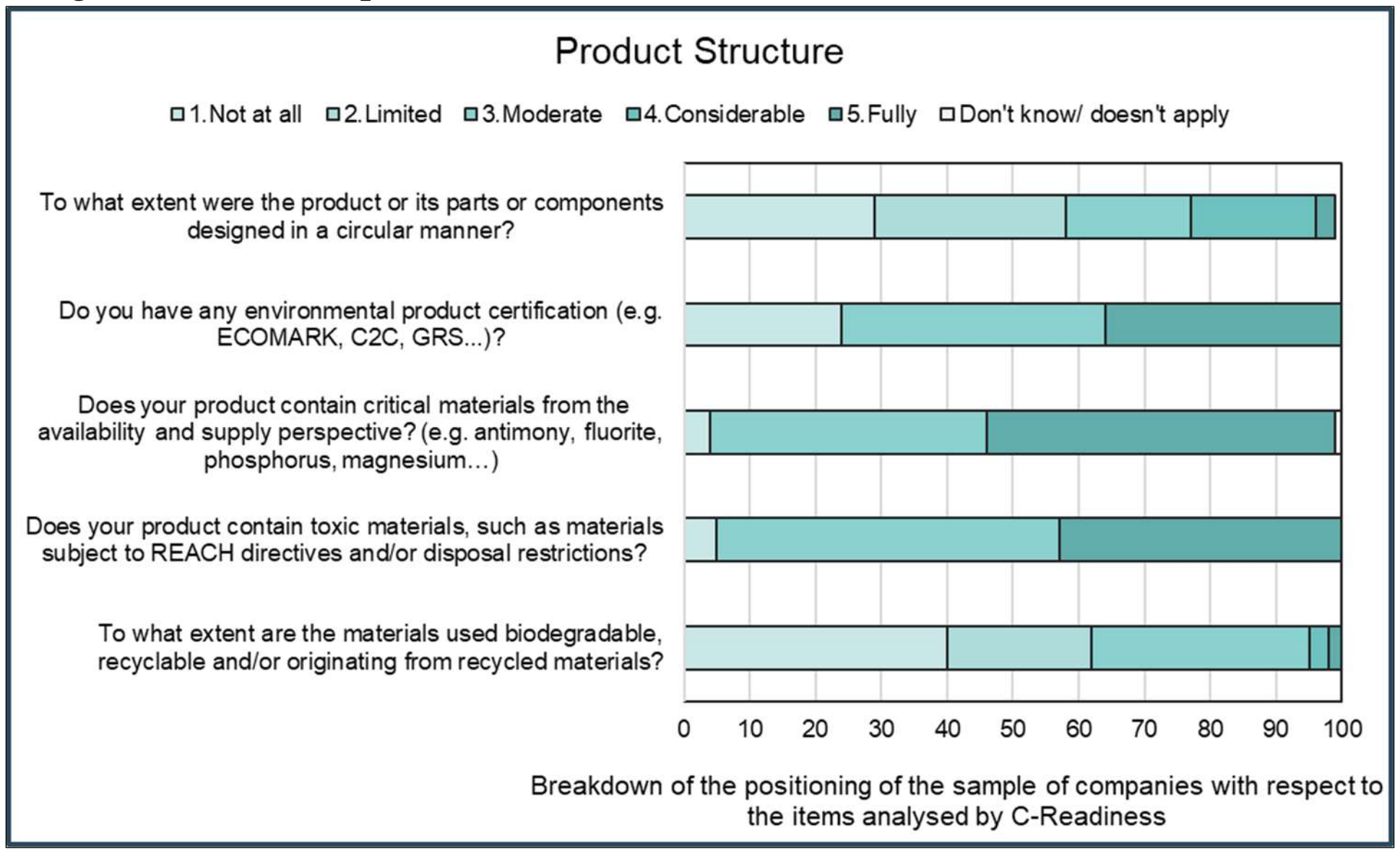

Figure A2: Categorization of Responses Based on Production Processes

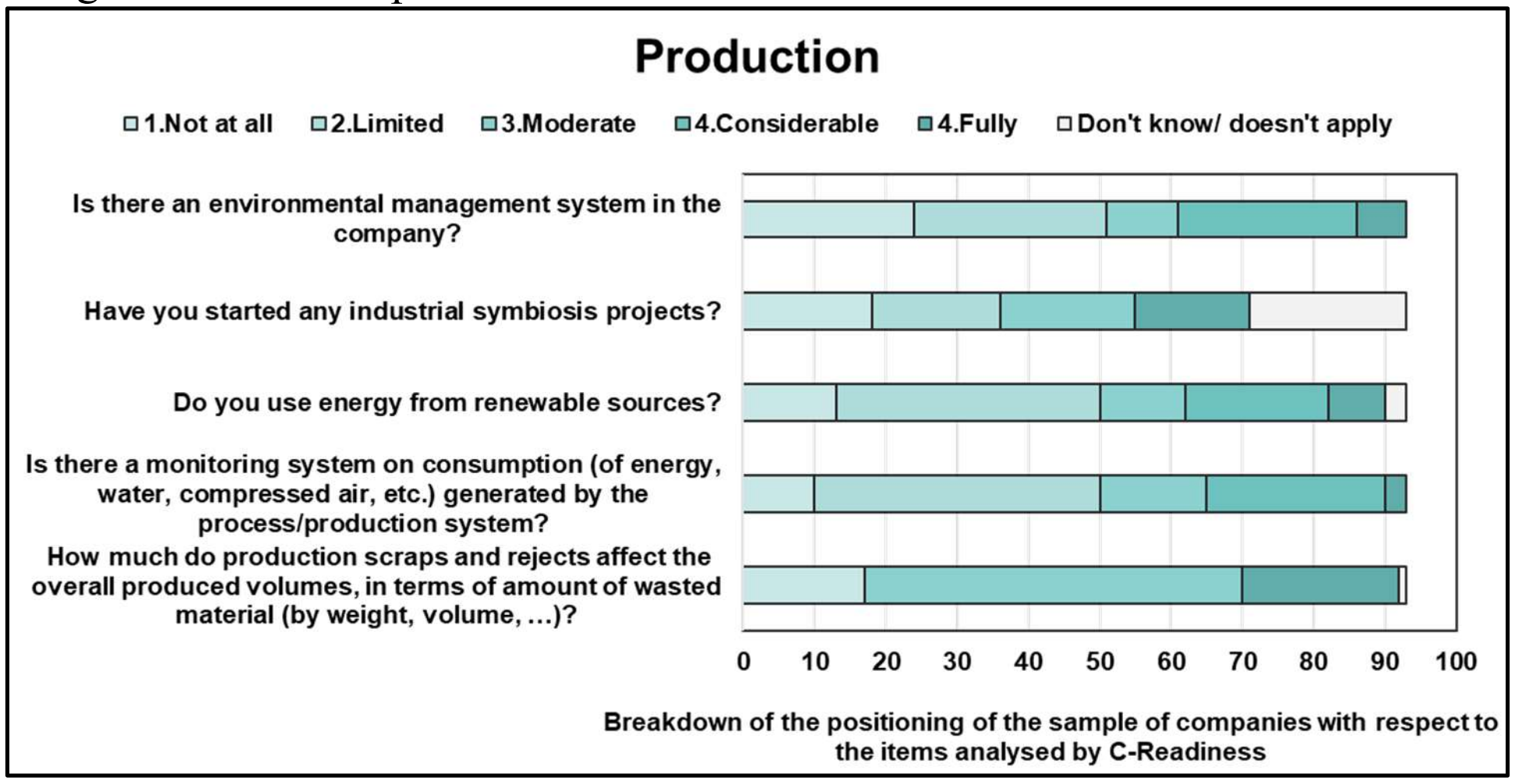

Figure A3: Categorization of Responses Based on Business Model

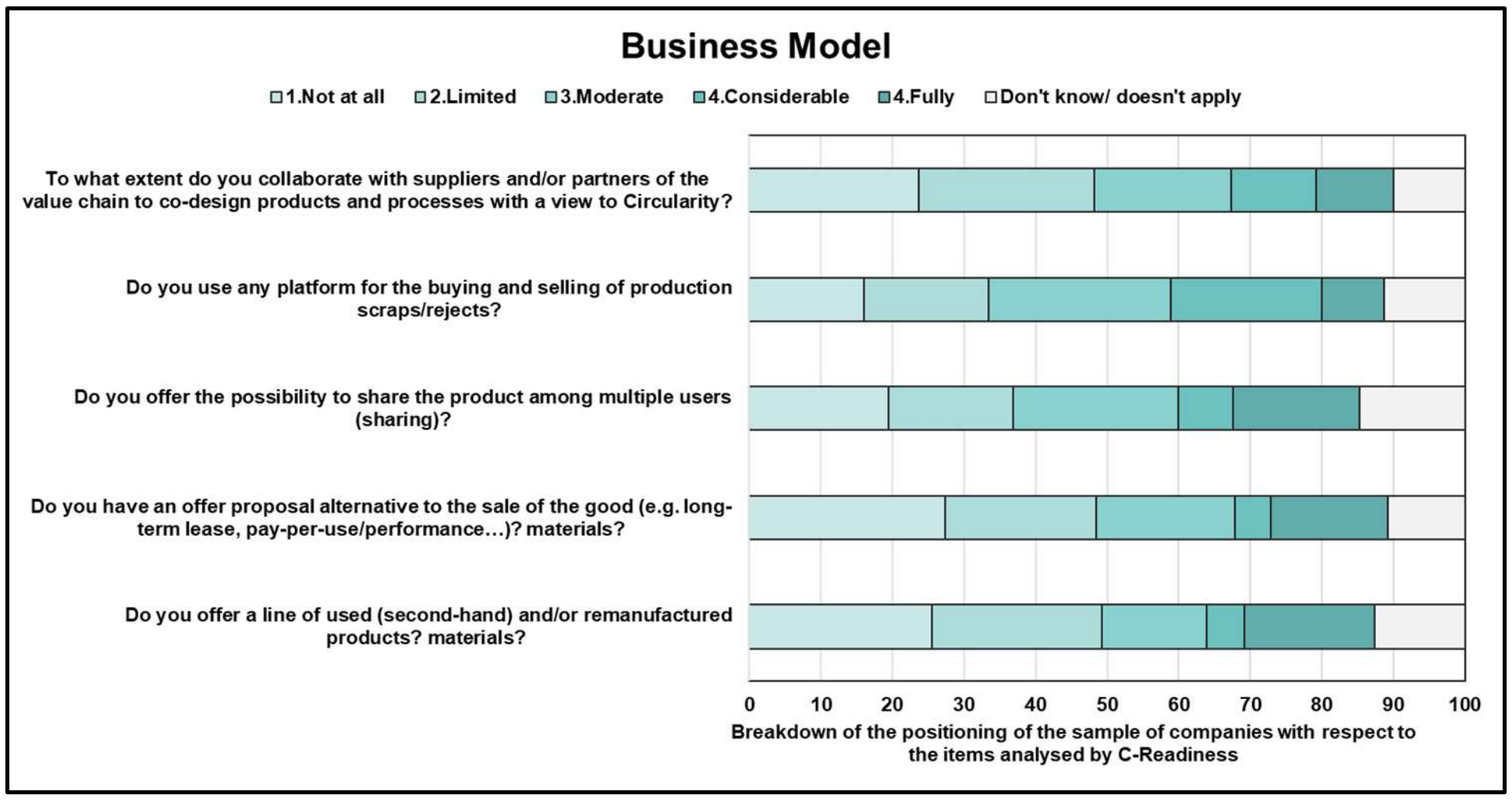

Figure A4: Categorization of Responses Based on Supply chain

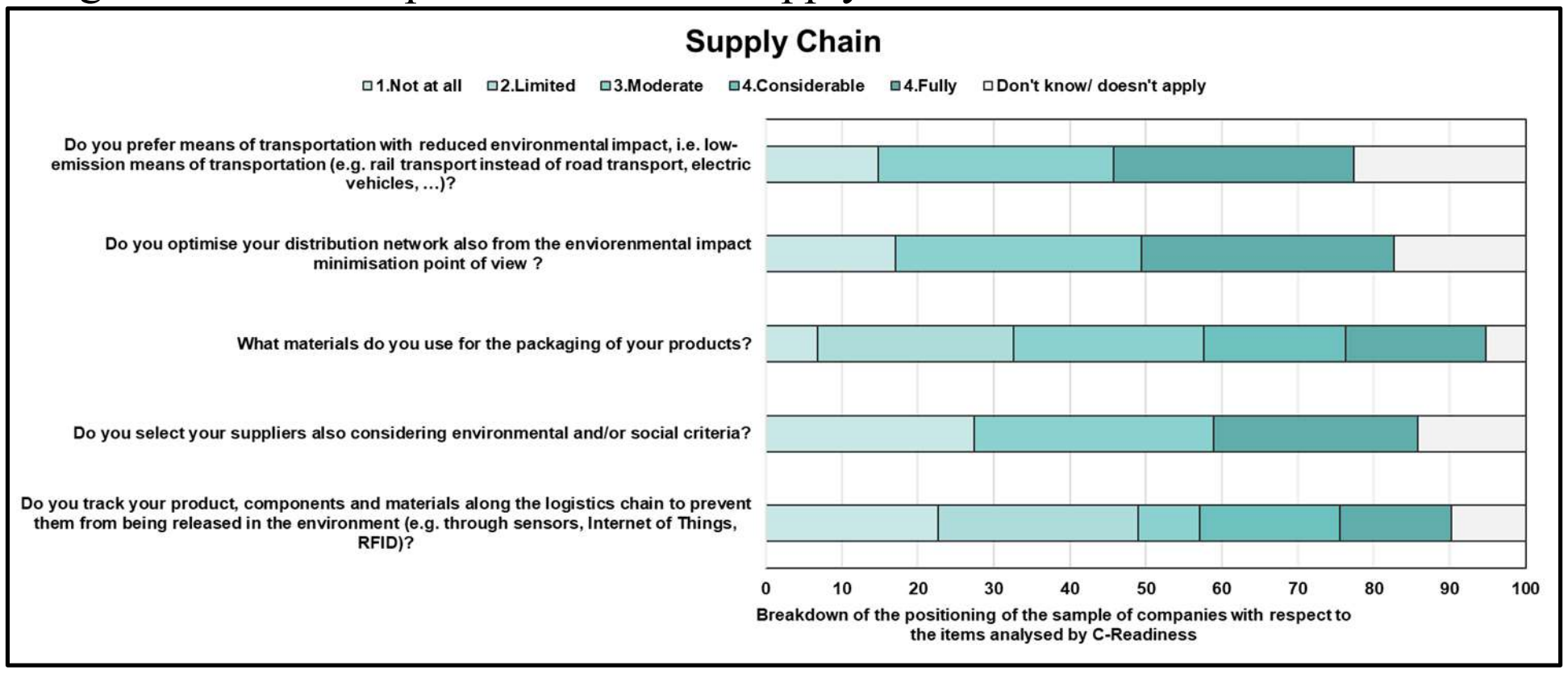

Figure A5: Categorization of Responses Based on Regeneration and End of life

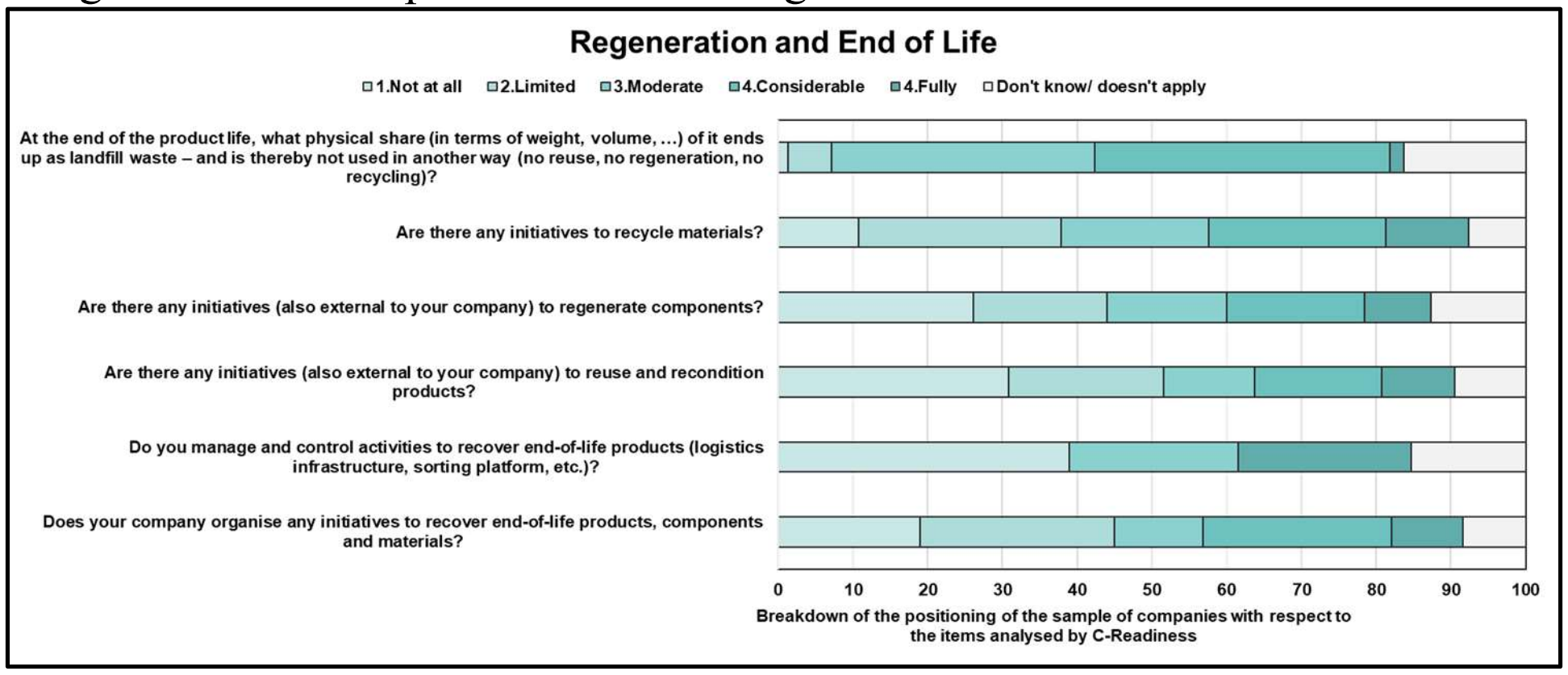

Figure A6: Categorization of Responses Based on Company Culture

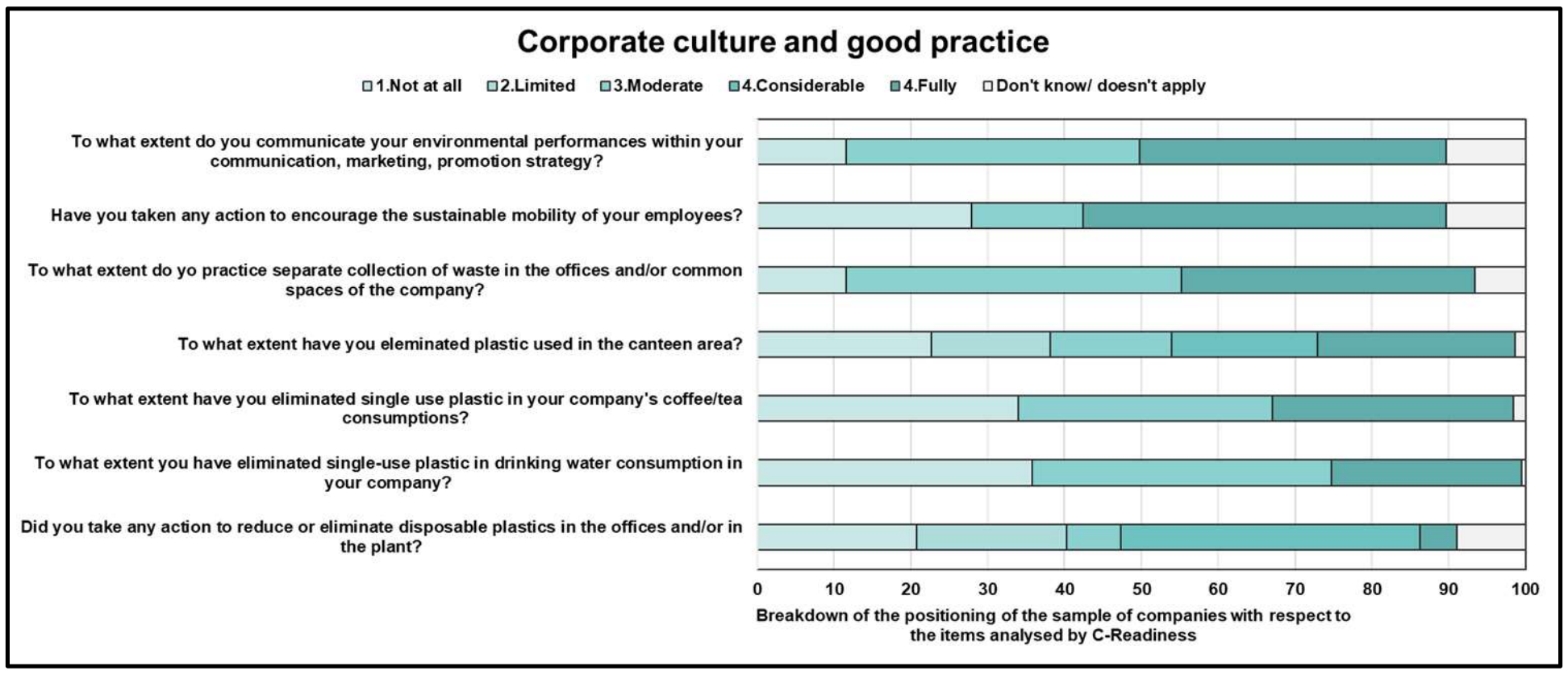

Corporate culture and good practice
1.Not at all
2.Limited
3.Moderate
4.Considerable
4.Fully
Don't know/ doesn't apply
To what extent do you communicate your environmental performances within your communication, marketing, promotion strategy?
Have you taken any action to encourage the sustainable mobility of your employees?
To what extent do yo practice separate collection of waste in the offices and/or common spaces of the company?
To what extent have you eleminated plastic used in the canteen area?
To what extent have you eliminated single use plastic in your company's coffee/tea consumptions?
To what extent you have eliminated single-use plastic in drinking water consumption in your company?
Did you take any action to reduce or eliminate disposable plastics in the offices and/or in the plant?
0
10
20
30
40
50
60
70
80
90
100
Breakdown of the positioning of the sample of companies with respect to the items analysed by C-Readiness

Appendix B1: Table for the Barries

| Categories | Barriers | Description | References |
|---|---|---|---|
| Financial or economic Barriers | Lack of high capital investment (B1) | Significant upfront costs and misaligned cost-revenue structures, the decoupling of costs and revenues to implement circular economy | (Bag et al., 2022; Bhattacharya & Kalakbandi, 2023; Chhimwal et al., 2022; Dutta et al., 2021; Gedam et al., 2021; Mangla et al., 2018; Sharma et al., 2024; V. S. Yadav & Majumdar, 2024; Zaidi & Chandra, 2024) |
| | Lack of financial support for CE adoption (B2) | Limited financial support and assistance for implementing a circular economy model | (Bag et al., 2022; Bhattacharya & Kalakbandi, 2023; Dutta et al., 2021; Jaiswal & Mukti, 2024; Mangla et al., 2018; Manoharan et al., 2022; Sharma et al., 2024; V. S. Yadav & Majumdar, 2024; Zaidi & Chandra, 2024) |
| | Lack of competitive virgin material pricing (B3) | The economic challenge posed by the lower cost of virgin materials compared to recycled or eco-friendly alternatives | (Chhimwal et al., 2022; Dutta et al., 2021; V. S. Yadav & Majumdar, 2024; Zaidi & Chandra, 2024) |
| | Perceived low returns and financial risks (B4) | This barrier refers to the apprehension among businesses that the initial investments in circular economy practices will not yield sufficient financial returns, leading to reluctance in adopting sustainable initiatives due to uncertainties in market demand and economic viability. | (Bag et al., 2022; Bhattacharya & Kalakbandi, 2023; Chhimwal et al., 2022; Dutta et al., 2021; Khan et al., 2022; Manoharan et al., 2022; Mhatre et al., 2023; R. Mishra et al., 2022; Sharma et al., 2024; V. S. Yadav & Majumdar, 2024; Zaidi & Chandra, 2024) |
| | Lack of access to affordable green technology (B5) | The difficulty businesses face in obtaining cost-effective and accessible sustainable technologies, which hinders their ability to implement circular economy practices | (Gedam et al., 2021; Jaiswal & Mukti, 2024; Mangla et al., 2018; Manoharan et al., 2022) |
| Operational Barriers | Lack of product design strategies for reuse and remanufacturing(B6) | The absence of effective design approaches that prioritize the ease of reuse and remanufacturing, limiting the potential for products to be recycled or repurposed at the end of their life cycle. | (Gedam et al., 2021; Jaiswal & Mukti, 2024; Mangla et al., 2018; Manoharan et al., 2022; Sharma et al., 2024; Zaidi & Chandra, 2024) |
| | Product complexity (B7) | The design complexity of products, characterized by the inclusion of toxic materials, rare materials and non-modular structures, poses significant barriers to effectively managing recovery processes within a circular economy framework. | (Sharma et al., 2023; V. S. Yadav & Majumdar, 2024; Zaidi & Chandra, 2024) |
| | Lack of design strategies for energy consumption reduction (B8) | To produce sustainable products and reduce energy consumption, many organizations are adopting altered design procedures, which can introduce complexity into the manufacturing process | (Bag et al., 2022; Dutta et al., 2021; Jaiswal & Mukti, 2024; Sharma et al., 2023; V. S. Yadav & Majumdar, 2024; Zaidi & Chandra, 2024) |
| Organisational Barriers | Lack of effective employee training and engagement for circular economy (B9) | Inadequate employee training and engagement can limit the potential of circular economy initiatives, skilled workforce is crucial for understanding and applying circular principles | (Bag et al., 2022; Bhattacharya & Kalakbandi, 2023; Dutta et al., 2021; Jaiswal & Mukti, 2024; Mangla et al., 2018; Sharma et al., 2023; V. S. Yadav & Majumdar, 2024; Zaidi & Chandra, 2024) |
| | Lack of cultural alignment with circular economy principles (B10) | limited awareness and commitment to the circular economy's principles among employees and managers. The hesitancy to adopt new, more sustainable practices is often due to a lack of understanding of the benefits or an organizational culture. | (Bag et al., 2022; Bhattacharya & Kalakbandi, 2023; Chhimwal et al., 2022; Dutta et al., 2021; Mangla et |
| | | | al., 2018; Manoharan et al., 2022; R. Mishra et al., 2022; V. S. Yadav & Majumdar, 2024; Zaidi & Chandra, 2024) |
| | Lack of top management motivation and resistance to change (B11) | Top management's insufficient expertise, lack of commitment, and resistance to change impede the adoption of circular economy practices. | (Chhimwal et al., 2022; Khan et al., 2022; Manoharan et al., 2022; R. Mishra et al., 2022; Sharma et al., 2023, 2024; V. S. Yadav & Majumdar, 2024; Zaidi & Chandra, 2024) |
| | Lack of industrial support and industrial symbiosis (B12) | Insufficient support from external stakeholders, such as technical experts and environmental agencies to implement industrial symbiosis | (Bag et al., 2022; Khan et al., 2022; V. S. Yadav & Majumdar, 2024; Zaidi & Chandra, 2024) |
| Market Barriers | Lack of a business plan for financial returns (B13) | Companies struggle to develop strategies that ensure profitability from returned products and optimize product portfolios. | (Bag et al., 2022; Bhattacharya & Kalakbandi, 2023; Chhimwal et al., 2022; Gedam et al., 2021; Jaiswal & Mukti, 2024; Khan et al., 2022; V. S. Yadav & Majumdar, 2024; Zaidi & Chandra, 2024) |
| | Lack of strategies to address product cannibalization (B14) | Circular products, like recycled or durable goods, can negatively impact the sales of existing products, leading to potential revenue loss. | (Zaidi & Chandra, 2024) |
| | Lack of access to know-how and expertise (B15) | Outsourcing circular activities, such as remanufacturing, can lead to a loss of control over product knowledge and intellectual property. | (Jaiswal & Mukti, 2024; R. Mishra et al., 2022; Zaidi & Chandra, 2024) |
| | Lack of market availability for reuse products (B16) | Limited consumer demand and market acceptance for reuse, remanufactured or refurbished products hinder their widespread adoption. | (Dutta et al., 2021; Jaiswal & Mukti, 2024; R. Mishra et al., 2022; S. Mishra et al., 2024; Sharma et al., 2024; Zaidi & Chandra, 2024) |
| Technological barriers | Lack of advanced technology and innovation (B17) | insufficient access to cutting-edge technologies, energy-efficient technologies, and upcycling techniques or lack of innovative solutions can impede the implementation of circular economy strategies. | (Badhotiya et al., 2022; Bhattacharya & Kalakbandi, 2023; Dutta et al., 2021; Jaiswal & Mukti, 2024; Khan et al., 2022; Mangla et al., 2018; R. Mishra et al., 2022; S. Mishra et al., 2024; Sharma et al., 2024; Zaidi & Chandra, 2024) |
| | Lack of data privacy and security measures (B18) | Inadequate protective protocols and regulations may hamper the safeguarding of privacy and data security in the collection of intelligent products at the end-of-use | (Bag et al., 2022; Bhattacharya & Kalakbandi, 2023; Dutta et al., 2021; Jaiswal & Mukti, 2024; Sharma et al., 2024; V. S. Yadav & Majumdar, 2024; Zaidi & Chandra, 2024) |
| | Lack of traceability in supply chains (B19) | Insufficient tracking of products throughout the supply chain hinders effective collection and recovery processes. | (Bag et al., 2022; Bhattacharya & Kalakbandi, 2023; Chhimwal et al., 2022; Dutta et al., 2021; Jaiswal & Mukti, 2024; Mhatre et al., 2023; V. S. Yadav & Majumdar, 2024; Zaidi & Chandra, 2024) |
| | Lack of infrastructure and resource capacity (B20) | Limited resources and outdated technologies impede the adoption of circular economy practices. | (Bag et al., 2022; Bhattacharya & Kalakbandi, 2023; Chhimwal et al., 2022; Gedam et al., 2021; Jaiswal & Mukti, 2024; Mhatre et al., 2023; V. S. Yadav & Majumdar, 2024; Zaidi & Chandra, 2024) |
| Environmental Barriers | Lack of awareness regarding environmental impacts (B21) | Limited understanding of environmental impacts, lack of awareness of life cycle assessment and industrial symbiosis hinder the adoption of sustainable practices. | (Bag et al., 2022; Bhattacharya & Kalakbandi, 2023; Chhimwal et al., 2022; Gedam et al., 2021; Jaiswal & Mukti, 2024; Mhatre et al., 2023; S. Mishra et al., 2024; Parida et al., 2023; Sharma et al., 2024; V. S. Yadav & Majumdar, 2024; Zaidi & Chandra, 2024) |
| | Unorganised recycling industry (B21) | The informal and disorganized nature of the recycling industry limits the effective collection and recycling of materials. | (Badhotiya et al., 2022; Bag et al., 2022; Bhattacharya & Kalakbandi, 2023; Jaiswal & Mukti, 2024; Khan et al., 2022; Manoharan et al., 2022; Parida et al., 2023; V. S. Yadav & Majumdar, 2024; Zaidi & Chandra, 2024) |
| | Lack of an appropriate system for reverse logistics (B22) | Inefficient reverse logistics systems hinder the ability to recycle products and achieve sustainability goals. | (Badhotiya et al., 2022; Dutta et al., 2021; Mangla et al., 2018; R. Mishra et al., 2022; Parida et al., 2023; V. S. Yadav & Majumdar, 2024) |
| Regulatory Barriers | Lack of taxation and incentives aligned with circular economy goals (B23) | The insufficient fiscal policies and incentives that fail to encourage businesses to adopt circular economy practices | (Badhotiya et al., 2022; Bag et al., 2022; Bhattacharya & Kalakbandi, 2023; Dutta et al., 2021; Gedam et al., 2021; Khan et al., 2022; Mangla et al., 2018; R. Mishra et al., 2022; S. Mishra et al., 2024; Parida et al., 2023; Sharma et al., 2023, 2024; V. S. Yadav & Majumdar, 2024; Zaidi & Chandra, 2024) |
| | Lack of policies, laws, and standard systems for CE (B24) | Complex governmental structures, fragmented policy frameworks, a lack of standards, and misaligned policies hinder the adoption of circular economy principles. | (Bag et al., 2022; Manoharan et al., 2022; Sharma et al., 2023, 2024; V. S. Yadav & Majumdar, 2024; Zaidi & Chandra, 2024) |
| | Lack of enforcement ability for existing legislation (B25) | The ineffectiveness in enforcing existing environmental regulations undermines efforts to transition towards CE practices | (Bag et al., 2022; Bhattacharya & Kalakbandi, 2023; Chhimwal et al., 2022; Mangla et al., 2018; Manoharan et al., 2022; Sharma et al., 2023, 2024; V. S. Yadav & Majumdar, 2024; Zaidi & Chandra, 2024) |
| | Lack of alignment between measures, metrics, and CE indicators (B26) | Current indicators, measures, and metrics fail to align with CE principles, making it difficult to evaluate progress accurately or drive improvements in resource efficiency, waste reduction, and sustainability efforts. | (Bag et al., 2022; Chhimwal et al., 2022; Dutta et al., 2021; Manoharan et al., 2022; Mhatre et al., 2023; S. Mishra et al., 2024; Sharma et al., 2023; V. S. Yadav & Majumdar, 2024; Zaidi & Chandra, 2024) |
| Supply chain (SC) management barriers | Lack of effective communication with suppliers (B27) | Poor communication and collaboration with suppliers can lead to delays, inefficiencies, and difficulties in implementing circular economy initiatives. | (Badhotiya et al., 2022; Bhattacharya & Kalakbandi, 2023; Gedam et al., 2021; Khan et al., 2022; Mangla et al., 2018; Manoharan et al., 2022; Zaidi & Chandra, 2024) |
| | Lack of coordination and information sharing among stakeholders (B28) | Insufficient coordination and information sharing among supply chain partners can hinder the adoption of circular economy practices | (Badhotiya et al., 2022; Bag et al., 2022; Bhattacharya & Kalakbandi, 2023; Chhimwal et al., 2022; Dutta et al., 2021; Gedam et al., 2021; Jaiswal & Mukti, 2024; Khan et al., 2022; Mangla et al., 2018; Manoharan et al., 2022; Mhatre et al., 2023; R. Mishra et al., 2022; S. Mishra et al., 2024; Parida et al., 2023; Sharma et al., 2023, 2024; V. S. Yadav & Majumdar, 2024; Zaidi & Chandra, 2024) |
| | Lack of effective supplier selection strategies (B29) | Selecting suppliers that do not align with circular economy principles can limit the potential for sustainable practices. | (Badhotiya et al., 2022; Bhattacharya & Kalakbandi, 2023; Gedam et al., 2021; Khan et al., 2022; Mangla et al., 2018; Manoharan et al., 2022; Zaidi & Chandra, 2024) |
| | Lack of supply chain designs for reuse and remanufacturing (B30) | Companies often struggle to optimize the design of forward-reverse logistics networks, which is essential for enabling the reuse and remanufacturing of products. | (Badhotiya et al., 2022; Bhattacharya & Kalakbandi, 2023; Gedam et al., 2021; Jaiswal & Mukti, 2024; Khan et al., 2022; Mangla et al., 2018; Manoharan et al., 2022; Mhatre et al., 2023; Parida et al., 2023; Sharma et al., 2024; V. S. Yadav & Majumdar, 2024; Zaidi & Chandra, 2024) |
| Cultural Barriers/Social Barriers /User Behaviour challenges | Lack of awareness and knowledge among consumers and suppliers (B31) | Insufficient awareness and knowledge among consumers and suppliers, along with the unavailability of qualified professionals, hinder the adoption of environmental management practices and the promotion of CE principles. | (Bag et al., 2022; Bhattacharya & Kalakbandi, 2023; Chhimwal et al., 2022; Dutta et al., 2021; Gedam et al., 2021; Jaiswal & Mukti, 2024; Manoharan et al., 2022; Mhatre et al., 2023; S. Mishra et al., 2024; Parida et al., 2023; Sharma et al., 2023, 2024; V. S. Yadav & Majumdar, 2024; Zaidi & Chandra, 2024) |
| | Lack of positive consumer perception towards circular economy practices (B31) | Consumers often perceive CE products as having compromised quality, creating a significant barrier to the acceptance and widespread adoption of sustainable goods and services. | (Badhotiya et al., 2022; Mhatre et al., 2023; V. S. Yadav & Majumdar, 2024; Zaidi & Chandra, 2024) |
| | Lack of strategies to address the loss of ownership value (B32) | CE practices, particularly value servitization models, may lead to a perceived or actual loss of ownership value, deterring businesses and consumers from fully embracing circular solutions. | (Zaidi & Chandra, 2024) |
| | Lack of encouragement for design for disassembly and recycling (B33) | Insufficient implementation of efficient take-back systems prevents the recovery of end-of-life products, limiting opportunities for remanufacturing and recycling and ultimately delaying the adoption of circular design principles. | (Badhotiya et al., 2022; Manoharan et al., 2022; R. Mishra et al., 2022; Sharma et al., 2023, 2024; Zaidi & Chandra, 2024) |
| | Lack of measures to mitigate circular rebound risks (B34) | The transition to a circular economy can lead to unintended consequences, such as increased resource consumption or pollution, if not carefully managed. | (Mhatre et al., 2023; S. Mishra et al., 2024; Zaidi & Chandra, 2024) |